\documentclass{article}
\usepackage{spconf,amsmath,graphicx}

\usepackage{floatflt}

\usepackage{color}
\usepackage{cite}

\usepackage{comment}

\usepackage{graphicx}
\usepackage{amsmath}
\usepackage{amsthm}
\usepackage{amssymb}

\usepackage{verbatim}

\usepackage{psfrag,array}

\usepackage{subfigure}

\usepackage{stfloats}

\usepackage{todonotes}

\usepackage{amsmath}
\usepackage{epstopdf}
\usepackage{algorithmic}

\newcommand{\p}[1]{\mathop{\mbox{\it p} } }

\newcommand{\be}{\begin{equation}}
\newcommand{\ee}{\end{equation}}
\newcommand{\ba}{\begin{array}}
\newcommand{\ea}{\end{array}}
\newcommand{\bea}{\begin{eqnarray}}
\newcommand{\eea}{\end{eqnarray}}
\newcommand{\bean}{\begin{eqnarray*}}
\newcommand{\eean}{\end{eqnarray*}}

\definecolor{white}{rgb}{1,1,1}

\begin{document}

\title{Spherical Large Intelligent Surfaces}

\name{Sha Hu$^{\dagger}$  \qquad Fredrik Rusek$^{\ddagger}$}

\address{$^{\dagger}$ Huawei Lund Research Center, Lund, Sweden \\
 $^{\ddagger}$Department of Electrical and Information Technology, Lund University, Lund, Sweden}

\maketitle

\begin{abstract}
As an emerging technology and evolution that goes beyond massive multi-input multi-output (MIMO), large intelligent surface (LIS) has gained much interest recently. LIS acts as an electromagnetic surface and can transmit, redirect, and receive radiating signals across its entire contiguous surface. It allows for unprecedented energy-focusing, data-transmission and terminal-positioning, and can fulfill the most grand visions for future communication systems. Earlier proposed LISs are in two-dimensional (2D), i.e., planar shapes. In this paper, we extend LIS to be three-dimensional (3D) and deployed as spherical surfaces. Compared to 2D shapes, spherical LISs have advantages of wide coverage, simple positioning techniques, and flexible deployments as reflectors.  
\end{abstract}

\begin{keywords}
Large intelligent surface (LIS), spherical surface, received signal strength (RSS), positioning, reflector.
\end{keywords}

\section{Introduction}
Large intelligent surface (LIS) is a promising technology for future wireless communication systems envisioned in \cite{HRE171, HRE181,  HRE182, HRE184}, which is beyond massive multi-input multi-output (MIMO) \cite{M10, MM12} and breaks a traditional antenna-array concept. LIS allows for exceptional focusing of energy in three-dimensional (3D) space, remote sensing with extreme precision, and unprecedented data-transmission. The abundant signal dimensions provided by LIS also facilitates potential applications powered by artificial intelligence (AI). 

Earlier proposed LISs are two-dimensional (2D) such as with squared and disk shapes. In its fundamental form, a LIS uses the entire contiguous surface for transmitting and receiving radiating signals. Under ideal assumptions, fundamental limits on the number of independent signal dimensions provided by a LIS per deployed surface-area are derived in \cite{HRE181}, where we show that the interference between two users is close to a \textit{sinc}-function. In \cite{HRE182}, Cram\'er-Rao lower-bounds (CRLB) for terminal-positioning with LIS are derived, which can decrease in the third-order of surface-area in near-field.

Implementing LIS in practice is however, challenging and requires new technologies in many fields such as physical material, radio-frequency (RF), and hardware integration. One observation form \cite{HRE181} is that after applying a match-filtering (MF), the received signal on LIS yields a band-limit property, and it can be implemented in a discrete-form with optimal sampling. Even so, a LIS still has a much larger scale of antenna-elements than a traditional massive MIMO system. On the other hand, using LISs as passive reflectors has been considered later in \cite{HY18, WZ18}, which can be seen as an extension of LIS to reflect-array antennas \cite{B63}. 

In this paper, we extend the LIS concept by proposing a spherical design depicted in Fig. 1, which can cooperate with future networks comprising of devices from the ground up to the space linked by terrestrial cellular networks, unmanned aerial vehicle (UAV) networks, and satellite systems. A spherical LIS can be seen as an evolution of spherical antenna-array proposed in \cite{CS68}, and it has many advantages compared to its planar shapes due to the geometric symmetry of a sphere. For instance, it can cover a wider range with higher received signal strength (RSS). It also yields simple positioning techniques, and can be flexibly deployed as a reflecting surface.

\begin{figure}[t]
\begin{center}
\vspace{0mm}
\hspace*{0mm}
\scalebox{1.25}{\includegraphics{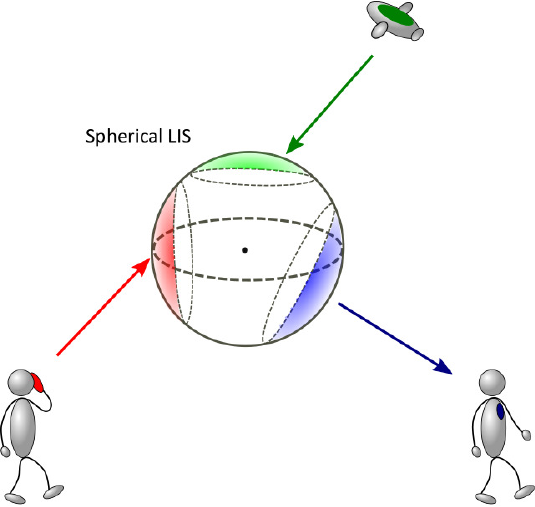}}
\vspace{-2mm}
\caption{\label{fig1}Three users communicating to a spherical LIS.}
\vspace{-8mm}
\end{center}
\end{figure}

\section{Spherical LIS}

\subsection{Received Signal Model}

The radiating signal from a terminal to a spherical LIS is depicted in Fig. 2. Expressed in Cartesian coordinates, the center of LIS is located at $x\!=\!y\!=\!z\!=\!0$, while the terminal is located at $(x_k,\,y_k,\,z_k)$. Due to the isotropic property of a sphere, we can equivalently let the terminal located at position $(0,\,0,\,\sqrt{x_k^2+y_k^2+z_k^2})$. Therefore, for simplicity we assume that the terminal is positioned at $(0,\,0,\,z_k)$, and for analytical tractability a perfect line-of-sight (LoS) propagation scenario is considered.

Following similar analysis in \cite{HRE171}, the received signal at location $(x,\,y,\,z)$ on the LIS (with a unit transmit-power and ignoring noise) can be modeled as
\bea \label{md1} s_k(x,y,z)=\sqrt{P_L\cos\psi}\exp\!\left(\!-2\pi j f_c \Delta_t\right), \eea
where the power attenuation is $ P_L\!=\!1/(4\pi \eta_k^2)$, and the elevation angle\footnote{Seen from Fig. 2, when the elevation angle is 0, the terminal is located upon the north pole of the sphere, while correspondingly when the angle is $\pi/2$, the terminal is located on the $xy$-plane.} $\psi$ denotes the angle-of-arrival (AoA). We assume a narrow-band system at a carrier-frequency $f_c$ whose wavelength is $\lambda$, and let $\Delta_t\!=\!\eta_k/(\lambda f_c)$ denote the transmit-time from the terminal to the LIS for a distance
\bea \eta_k=\sqrt{x^2+y^2+(z-z_k)^2}.\eea 

Seen from Fig. 2, it holds from cosine theorem that
\bea \cos\psi=\frac{z_k^2-\eta_k^2-R^2}{2R\eta_k},\eea
where $R$ is the radius of the sphere. Transforming to spherical coordinates by setting
 \bea x&\!\!\!=\!\!\!&R\sin\theta\cos\phi, \notag \\
  y&\!\!\!=\!\!\!&R\sin\theta\sin\phi,  \\
   z&\!\!\!=\!\!\!&R\cos\theta,  \notag
 \eea
and combining (1)-(4) yields
\be \label{md2} s_k(x,y,z)=\frac{1}{2\sqrt{\pi}}\frac{\sqrt{\tau\cos\theta-1}}{(1-2\tau\cos\theta+\tau^2)^{3/4}}\exp\!\left(\!-\frac{2\pi j \eta_k}{\lambda}\right). \ee
The distance $\eta_k$ in (2) changes to
\bea \eta_k=R\sqrt{1-2\tau\cos\theta+\tau^2}, \eea
by denoting a normalized distance as
\bea \tau=z_k/R.\eea

The surface-region where the LIS can receive (transmit) signals from (to) the terminal is spanned by the elevation angle-region $\theta\!\in\![0,\,\theta_0]$, and the maximal angle is
\bea \label{theta} \theta_0=\arccos(1/\tau),\;\,0\!\leq\!\theta_0\!\leq\!\pi/2. \eea

\subsection{Advantages of a Spherical LIS over a Planar LIS}

There are several advantages of a spherical LIS compared to a planar one. First if all, rotating a terminal around the center of sphere does not change the information theoretical properties. This brings substantial gains in average received signal strength (RSS) with terminal mobility. It can also have simple positioning methods, based on the boundary on the surface where the LIS can receive signal, or equivalently, measuring $\theta_0$ as in (\ref{theta}), which utilizes the geometric shape of the LIS. In ideal implementations, $\theta_0$ can be accurately estimated and the terminal can be positioned upon the north pole (after certain rotation). Another advantage is flexible designs as reflecting surfaces when partial of the spherical LIS is used for collecting signal from a base-station, and meanwhile the other part is used for redirecting the signal to a targeted terminal that is blocked out from the base-station. Moreover, the channel phases can be compensated both at receiving and transiting stages based on the estimated positions of the base-station and the terminal, respectively. These nice properties obtained from a spherical LIS  are explained in what follows.

\begin{figure}[t]
\begin{center}
\vspace{0mm}
\hspace{2mm}
\scalebox{.27}{\includegraphics{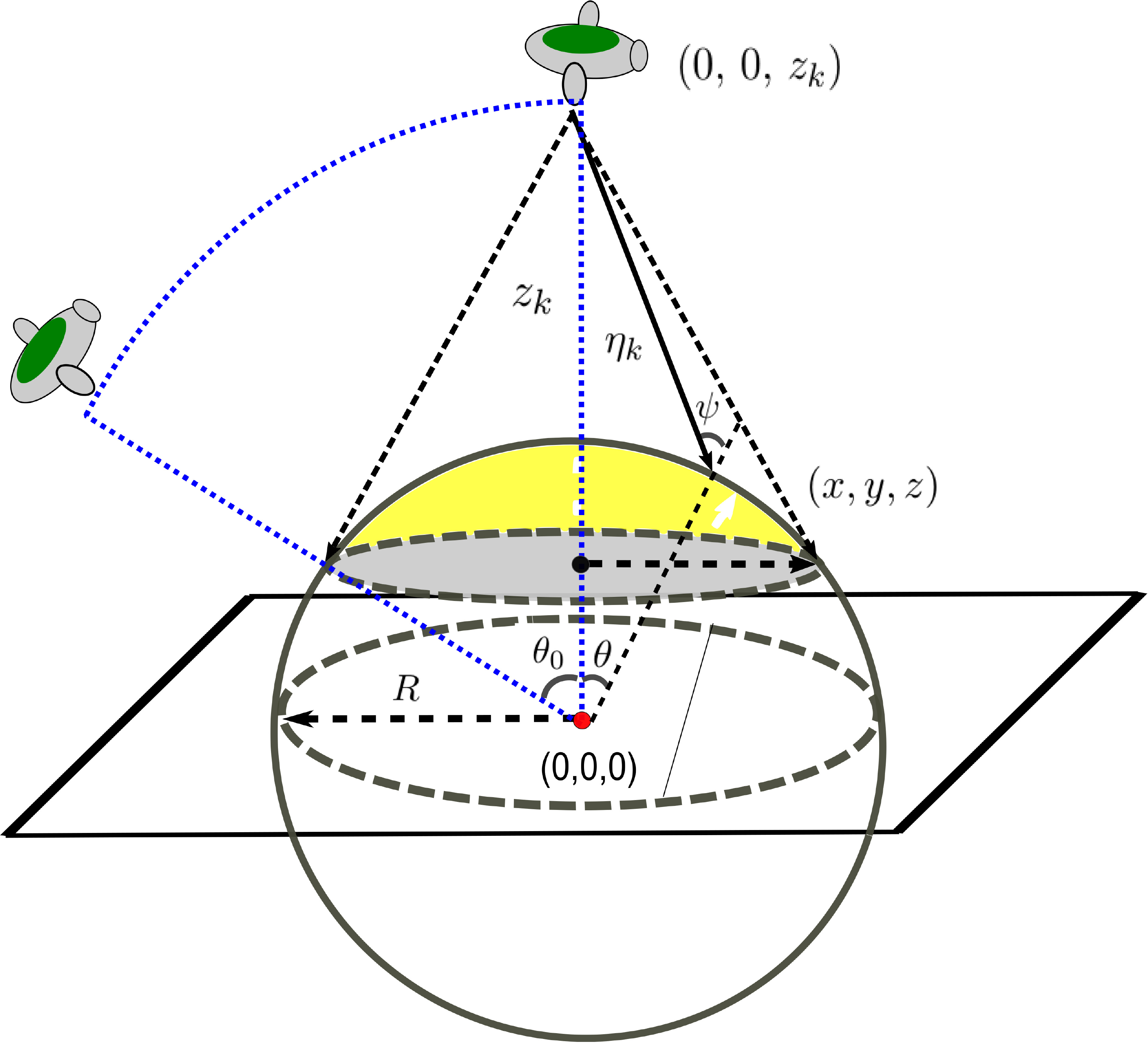}}
\vspace{-4mm}
\caption{\label{fig2}Radiating signal model with a spherical LIS.}
\vspace{-8mm}
\end{center}
\end{figure}

\subsection{Array Gains with a Spherical LIS}

We first compare the RSS between a spherical and a planar LIS. Note that the maximal effective surface-area of a spherical LIS that can receive signals from a terminal is half the surface-area, i.e., $2\pi R^2$ ($\theta_0\!=\!\pi/2$). We shall compare it to a planar LIS (assuming a disk-shape) with the same surface-area, i.e., with a radius of $\sqrt{2}R$. The RSS on a spherical LIS of a radius $R$, for a given $\tau$ and a maximal integral angle $\hat{\theta}$, is equal to (by noting that the Jacobian of transformation from Cartesian to spherical coordinates is equal to $R^2\sin\theta$)
\bea \label{rss0} P_{\text{sp}}( \tau, \hat{\theta})&\!\!\!\!=\!\!\!\!&\int_0^{2\pi} \!\!\! \int_{0}^{\hat{\theta}} |s_k|^2R^2\sin\theta\mathrm{d}\theta\mathrm{d}\phi \notag \\
&\!\!\!\!=\!\!\!\!&\!\frac{1}{2}  \int_0^{\hat{\theta}} \!\!\!\frac{(\tau\cos\theta-1)\sin\theta}{\big(1-2\tau\cos\theta+\tau^2\big)^{3/2}}\mathrm{d}\theta \notag\\ 
&\!\!\!\!=\!\!\!\!&\frac{1}{2}\left(\!1-\frac{\tau-\cos\hat{\theta}}{\sqrt{\tau^2-2\tau\cos\hat{\theta}+1}}\right)\!. \qquad \eea
When receiving at the maximum $\hat{\theta}\!=\!\theta_0$, it holds that
\bea \label{rss1} P_{\text{sp}}( \tau, \theta_0)=\frac{1}{2}\left(1-\frac{\sqrt{\tau^2-1}}{\tau}\right)\!.   \eea

On the other hand, the RSS of a planar LIS with a disk-shape and of a radius $R$, for a given terminal located at $(0,\,z_k\sin\theta,\,z_k\cos\theta)$ (for a disk-shaped LIS, one can rotate the disk such that the terminal is located on $x\!=\!0$), can be shown from \cite{HRE181} that
\bea \label{rss2} P_{\text{pl}}(\tau, \theta)&\!\!\!\!\!\!=\!\!\!\!\!\!&\int_0^{2\pi} \!\!\! \int_{0}^{R}r |s_k|^2\mathrm{d}r\mathrm{d}\phi \notag \\
&\!\!\!\!\!\!=\!\!\!\!\!\!&\frac{1}{4\pi}\!\int_0^{2\pi}\! \!\!\!\int_0^{R} \!\!\!\! \frac{r z_k\cos\theta}{\big(z_k^2-2rz_k \sin\theta \sin\phi+ r^2\big)^{3/2}}\mathrm{d}r\mathrm{d}\phi \notag \\
&\!\!\!\!\!\!\approx\!\!\!\!&\frac{z_k\cos\theta}{2}\!\int_0^{R} \!\!\! \frac{r }{\big(z_k^2+ r^2\big)^{3/2}}\mathrm{d}r  \notag \\
&\!\!\!\!\!\!=\!\!\!\!\!\!&\frac{\cos\theta}{2}\left(1-\frac{\tau}{\sqrt{\tau^2+1}}\right)\!. \qquad  \eea
The approximation holds when $\tau\!\gg\! 1$. When $\theta\!=\!0$, i.e., the terminal is located at $(0,\,0,\,z_k)$, the equality in (\ref{rss2}) is exact. 

Before proceeding any further, we first point out a fact that for a planar LIS, $\tau$ can be as small as possible. However, for a spherical LIS, the minimal $\tau$ is 1 when the terminal is infinitely close to the surface.

Note that when a terminal moves around the sphere with a fixed radius $z_k$, i.e., $\tau$, the RSS measured on the spherical LIS is unaltered. However, for a planar LIS, the RSS changes approximately by a factor of $\cos\theta$ seen from (\ref{rss2}) when $\tau\!\gg\! 1$. Averaging $\theta$ over $[0,\,\pi/2)$, the RSS ratio of a spherical LIS with a radius $R$ over a planar LIS with a radius $\sqrt{2}R$ is
\bea \gamma=\frac{ \frac{\pi}{2}P_{\text{sp}}(\tau, \theta_0)}{\int_0^{\pi/2}\text P_{\text{pl}}(\tau/\sqrt{2}, \theta)\mathrm{d}\theta}=\frac{\pi}{2} \left(\frac{1-\frac{\sqrt{\tau^2-1}}{\tau}}{1-\frac{\tau}{\sqrt{\tau^2+2}}}\right).   \eea

When $R$ is sufficiently large, it holds that $P_{\text{sp}}(1)\!=\!P_{\text{pl}}(0)=1/2$, since half of the transmitting power propagating towards the LIS are received both on a spherical and a planar LIS. Further, to see array gains with a spherical LIS compared to a planar one, we let $\tau\to \infty$ which yields
\bea \gamma=\pi/2.   \eea
On the other extreme, letting $\tau\to 1$ yields
\bea \gamma= \frac{\sqrt{3}\pi}{2(\sqrt{3}-1)}\approx 3.7.   \eea
However, when $\tau\to 1$ the approximation of $P_{\text{pl}}(\tau, \theta)$ in (11) becomes less accurate, and the limit may deviate from the true value with a small $\tau$. 

%
%
%

\subsection{Positioning with Spherical LIS}

With a spherical LIS, the positioning can become quite simple, since only $z$-dimension needs to be estimated. Considering a LoS scenario, based on the measured angle $\tilde{\theta}_0$ on the LIS, the direction of the terminal aligns with the line that connects the center of sphere and the center of the cross-section circle (specified by $\tilde{\theta}_0$ in Fig. 2) corresponding to the cone that can receive signal. Meanwhile, the distance is estimated from (\ref{theta}) as
\bea \tilde{z}_k=R/\cos\tilde{\theta}_0. \eea

Alternatively, we can also set 
\bea \theta_n=\tilde{\theta}_0/n. \eea
and measure a series of RSS, which holds from (\ref{rss0}) that
\bea  P_{\text{sp}}(\tau, \theta_n)=\frac{1}{2}\left(\!1-\frac{\tau-\cos\theta_n}{\sqrt{\tau^2-2\tau\cos\theta_n+1}}\right)\!. \eea
Then, one can obtain a series of estimates for $\tau$, and based on those the distance $z_k$ can be estimated.

Under Non-LoS (NLOS) scenario, there could be multiple paths reaching the LIS. Similar to positioning with traditional antenna-array systems, one needs to resolve the multiple-path components (MPC) and find the strongest LoS path with techniques such as successive interference cancellation (SIC) \cite{HRE173}. Further, if there are multiple LIS deployed around, estimates from different LISs can be combined to refine the positioning results.

Nevertheless, we consider the CLRB for positioning $\tau$ with a spherical LIS under LoS case. A direct approach would be to evaluate the Fisher-information \cite{HRE182} with the received signal model (\ref{md2}). As we are interested in the relationship between the CRLB and the radius $R$, we use RSS in (\ref{rss1}) and (\ref{rss2}) to compute the CRLB by assuming that the noise is additional white Gaussian noise (AWGN). 

Noting that
\bea \text{CRLB}_{\text{sp}}(\tau)\propto\left(\! \frac{\partial P_{\text{sp}}(\tau)}{\partial \tau}\!\right)^{-2} 
=4\tau^4(\tau^2-1), \eea
it shows that RSS based positioning of $\tau$, i.e., $z_k$ decreases in $R^{-6}$, or the third-order of surface-area. This shall be compared to the case with a planar LIS. 

From (\ref{rss2}) it holds that
\bea \text{CRLB}_{\text{pl}}(\tau)\propto\left(\! \frac{\partial P_{\text{pl}}(\tau)}{\partial \tau}\!\right)^{-2} 
=4(\tau^2+1)^3, \eea
which also decreases in $R^{-6}$ and aligned with \cite{HRE181}. But as expected, the CRLB with a spherical LIS is smaller than that with a planar LIS for a given $\tau$. Surprisingly, we see that with a spherical LIS the CRLB goes to zero when $\tau\to 1$. However, with a planar LIS the CRLB has a fundamental limit 4 as $\tau\to 0$. This also partly because that the approximation in (11) becomes less accurate when $\tau$ is small.

\subsection{Using Spherical LISs as Reflecting surfaces}

As depicted in Fig. 2, a spherical LIS can also be used as a reflecting surface for low-cost circuits and RF designs. In this case, the received signal from a traditional base-station can be collected using part of the surface towards it, and then redirected by using another part of surface facing a terminal that is blocked out to the base-station. Due to the simple positioning methods introduced earlier with a spherical LIS, both positions of the base-station and the terminal can be estimated. With those, the channel phases can be compensated when receiving from the base-station and also when reflecting to the terminal. This yields a very flexible and effective designs of using LISs as reflecting surfaces.

\begin{figure}[t]
\begin{center}
\vspace{-0mm}
\hspace{-5mm}
\scalebox{.34}{\includegraphics{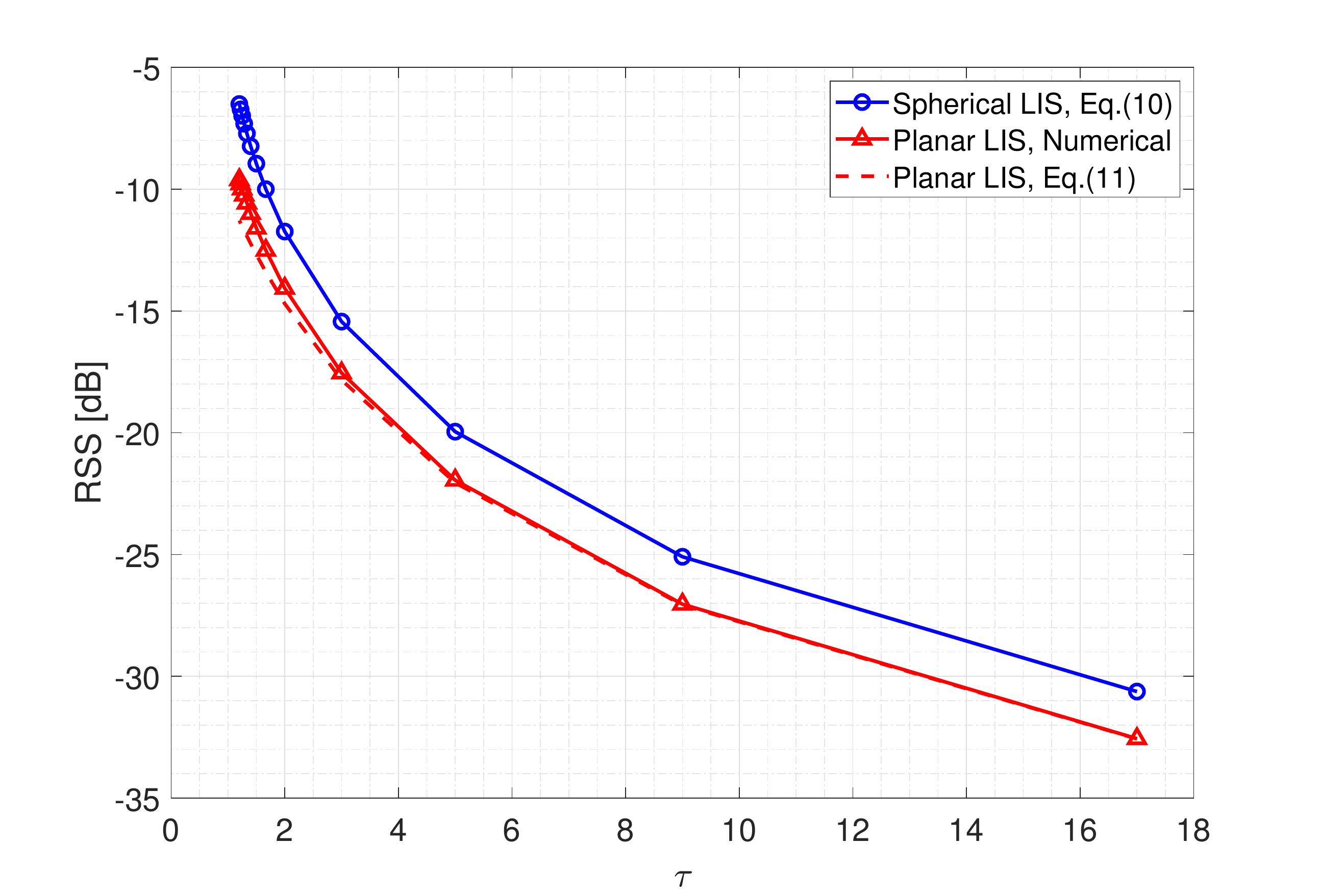}}
\vspace{-4mm}
\caption{\label{fig4}RSS with a spherical and a planar LIS with $z_k\!=\!4$.}
\vspace{-8mm}
\end{center}
\end{figure}

\subsection{Numerical Results}

Below we show some simulation results with spherical LISs compared to planar ones to verify our analysis. 

In Fig.~3, the averaged RSS are measured with $z_k\!=\!4$ (in meter), where the radius $R$ (and $\tau$) changes. As can be seen, when $\tau\!>\!2$, the approximation of RSS for the planar LIS in (\ref{rss2}) are quite accurate and aligned with numerical results. Further, spherical LIS provides substantial RSS gains. 

The ratio $\gamma$ is plotted in Fig.~4, where we can see that as $\tau\to\infty$, $\gamma$ converges to $\pi/2$ both for numerical and analytical results, which are well aligned with the analysis in (13). However, as $\tau<5$, the numerical values of $\gamma$ become smaller than the analytical ones in (14), due to approximation errors. Nevertheless, we see that RSS gains with a spherical LIS are substantial for all values of $\tau$.

In Fig.~5, we compare the CRLBs obtained in (18) and (19), respectively. When $R$ is small, the CRLBs are close and decrease in the sixth order of $R$. As $R$ increases (i.e., $1/\tau$ increases), the spherical LIS outperforms a planar LIS.

\begin{figure}
\begin{center}
\vspace{-0mm}
\hspace{2mm}
\scalebox{.34}{\includegraphics{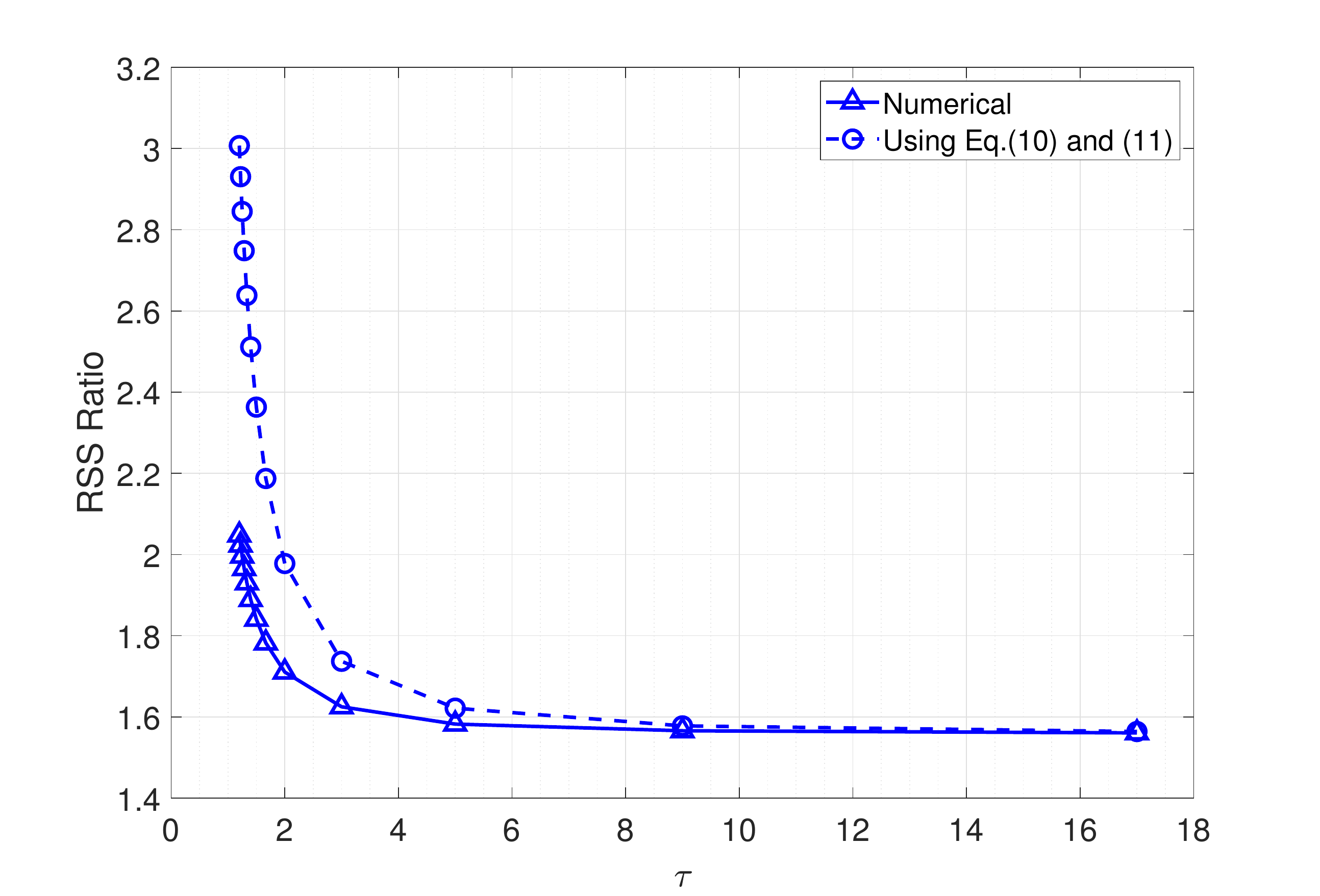}}
\vspace{-8mm}
\caption{\label{fig5}RSS ratio $\gamma$ with a spherical and a planar LIS with $z_k\!=\!4$.}
\vspace{-7mm}
\end{center}
\end{figure}

\begin{figure}[t]
\begin{center}
\vspace{-0mm}
\hspace{5mm}
\scalebox{.325}{\includegraphics{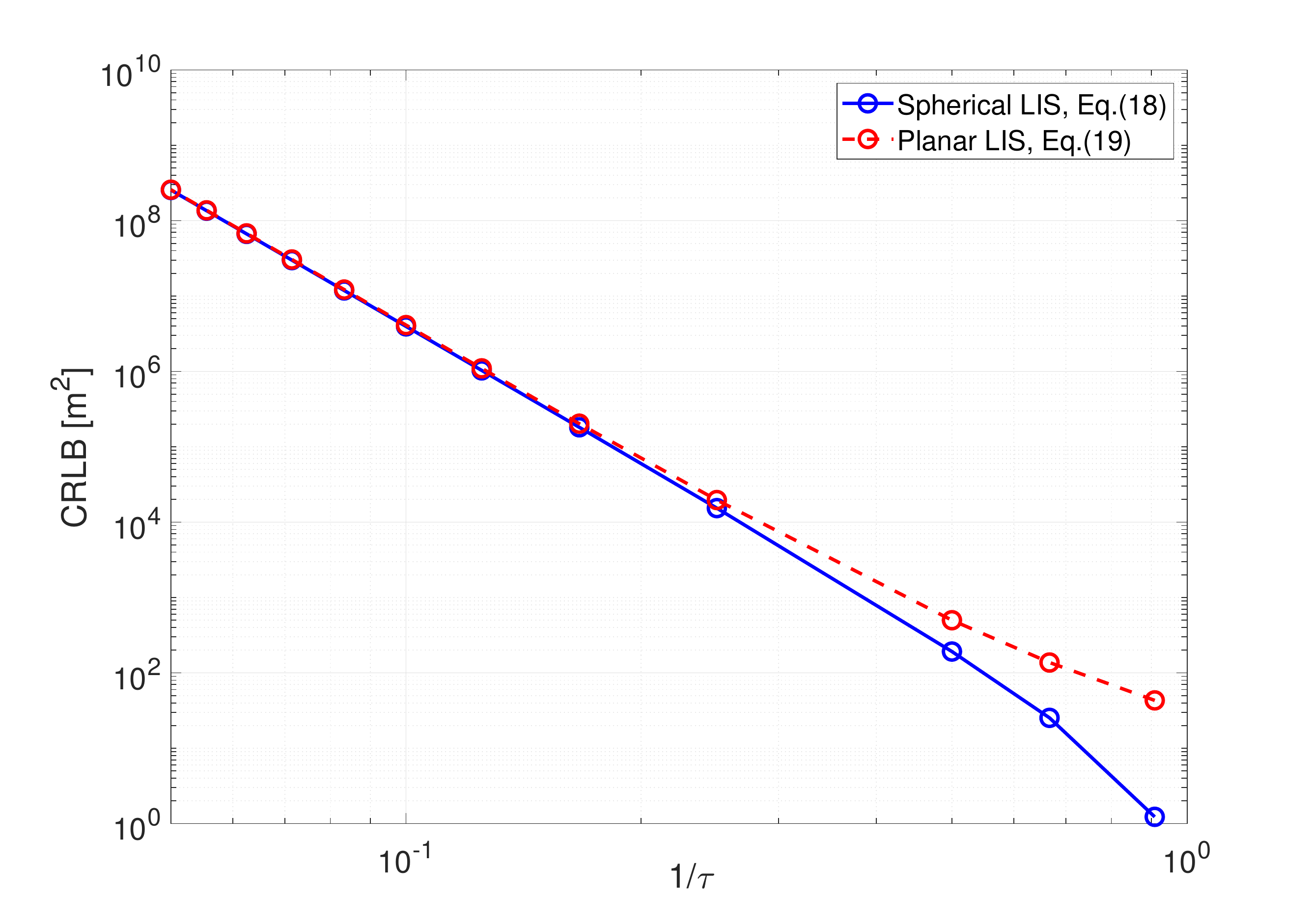}}
\vspace{-8mm}
\caption{\label{fig6}Positioning CRLB with a spherical and a planar LIS with $z_k\!=\!4$.}
\vspace{-8mm}
\end{center}
\end{figure}

\section{Summary}

In this paper, we have extended earlier proposed 2D planar LISs into 3D spherical designs. We have shown analytically that a spherical LIS can yield higher averaged RSS when a terminal moves around, and also lower positioning CRLB, compared to a planar LIS. Asymptotic properties are also analyzed which are aligned with simulation results. In addition, a spherical LIS can also be flexibly applied as low-cost controllable reflectors.

\clearpage

\bibliographystyle{IEEEtran}

\end{document}